**A bent straw as a tool for an affordable student-safe experiment in vortex ring dynamics**

Elijah James,[1] Yukun Sun,[1] Yicong Fu,[1] Jena Shields,[2] Cade Sbrocco,[3] Christopher Dougherty,[1] Chris Roh[1]

[1]Department of Biological and Environmental Engineering, Cornell University, New York, USA
[2]Department of Applied and Engineering Physics, Cornell University, New York, USA
[3]Department of Physics, Cornell University, New York, USA

**Abstract**

Vortex dynamics are an important topic in fluid dynamics, explaining phenomena like drag and lift generation, jet propulsion, and corner flows. It is also often excluded from introductory or undergraduate fluid dynamics courses on account of its complexity and the inaccessibility of practical and engaging experiments. We present an affordable student-safe experiment to generate vortex rings and study their dynamics using a bent straw and dyed water that allows students to control key parameters, can be imaged using a smartphone camera, and explains the complex physics with simple and easily measured parameters. Vortex rings are produced that parallel seminal experiments, demonstrating secondary structures and the mirroring effect. Meanwhile, nonplanar and triangular jet exits are used to demonstrate asymmetric vortex rings and vortex ring inversion.

## I. BACKGROUND

A vortex is a fundamental flow structure arising across vast length scales, spanning from centimeter scale breathing jets of aquatic insects to planetary flows in Jupiter.[1,2] These swirly motions carry important consequences in engineering and nature and are central to understanding complex phenomena such as turbulence, weather, and flight. However, teaching vortex dynamics is often reserved for advanced fluid dynamics courses; many students are left unaware of the rich and complex flow physics underlying processes like drag and lift generation, jet propulsion, and corner flows. The prevalence of vortex flows underscores their importance and should be included in undergraduate fluid dynamics curricula to give students an appreciation for the range of fundamental flow phenomena and guiding physics. This subject can also be used to visualize concepts from related curricula like electromagnetism.

Vortex dynamics provides opportunities to revisit engineering concepts such as solenoidal vector fields, Stokes' theorem and the Biot-Savart law in an analogous physical system. The Biot-Savart law presents a relation between a line vortex and its "induced" velocity: a small segment of the vortex line, $dl$, at position $\boldsymbol{x'}$ with vortex strength or circulation $\Gamma$ (defined as $\Gamma = \oint \boldsymbol{u} \cdot \boldsymbol{dx}$, where $\mathbf{u}$ is the velocity field) induces velocity $d\boldsymbol{u}$ at position $\mathbf{x}$ given by

$$d\boldsymbol{u}(\boldsymbol{x}) = -\frac{\Gamma}{4\pi}\frac{\boldsymbol{s} \times d\boldsymbol{l}(\boldsymbol{x}')}{s^3} \qquad (1)$$

where $\boldsymbol{s} = \boldsymbol{x} - \boldsymbol{x'}$ describes the relative position vector from the vortex line to a point of inquiry, while $s$ represent the magnitude of the relative position vector.[3] A simple way to visualize the Biot-Savart law is to use the "right-hand rule," where the direction of $\boldsymbol{s}$ crossed with the direction of the vortex line (***dl***) gives the direction of the induced velocity. Because some utility of the vorticity equation may be out of scope for introductory courses, new techniques to visually demonstrate some of the more intricate mathematical treatments are warranted.

Beyond theoretical treatment, interactive lecture demonstrations are effective at enhancing conceptual understanding.[4-6] One common tool for demonstrating vortex rings is a vortex cannon. The generator consists of a short and broad barrel with a slight taper closed by an elastic membrane on one side. After filling the barrel with smoke, visible coherent vortices can be created by stretching and releasing the elastic membrane. While this tool effectively demonstrates the presence and coherence of vortices, it lacks the ability to control key parameters, such as the strength of the vortex, which would allow students to navigate the underlying physics and promote active learning—especially as lecture demonstrations are often underutilized.[7-9]

The vortex strength, for example, is quantified using circulation, $\Gamma$. The mathematical expression of circulation can be interpreted as flow speed tangent to the closed path, which indicates a summation of the flow looping along the path. Using Stokes' theorem, circulation can also be expressed in the following way, $\Gamma = \int (\boldsymbol{\nabla} \times \boldsymbol{u}) \cdot \boldsymbol{n} dA$, where $\boldsymbol{n}$ is the outer normal unit vector and $\boldsymbol{\nabla} \times \boldsymbol{u}$ is curl of the velocity known as vorticity, $\boldsymbol{\omega}$. Accordingly, another way to interpret circulation is as the sum of vorticity inside the enclosing path. Vorticity can be interpreted as an angular velocity of a fluid parcel about its own center. Note, it is important to distinguish angular velocity from vorticity: fluid with angular velocity rotates about a common center whereas fluid parcels with vorticity rotate about their individual centers.

In the research laboratory, recent experiments in vortex dynamics have created rings with impulsive starting jets using a piston-cylinder system in a water tank, which allows for control of the amount of circulation in the vortex ring.[10-12] To initiate ring formation, a controlled valve determines the flow rate from a constant-pressure head tank, which then controls the piston velocity and the volume of displaced fluid moving through the jet exit. The fluid is marked with fluorescent dye and recorded to map the flow field and image ring development. A vortex ring is generated as the impulsively started jet injects vorticity into the quiescent fluid. As the piston moves, the ejected fluid rolls up into a ring and is continuously fed with vorticity such that it grows, separates, and propagates downstream.[10-12] Different nozzle geometries are also used to highlight different governing physics. For example, triangular jet exits demonstrate flip-flopping of the triangular ring, which exhibit the effect of "induced" velocity by a vortex, while rings generated with an inclined jet exit have shown rapid loss of coherence due to variation in circulation in primary vortex ring.[11,13]

In these experiments, vortex dynamics is described by the so-called *vortex formation time*, $\hat{T}$, defined as

$$\hat{T} = \frac{UT}{D} \qquad (2)$$

where $U$ is the jet velocity and $T$ is the duration of jet injection and $D$ is the length scale of the exit. Vortex formation time allows us to examine the evolution of a vortex ring by parsing dimensional time into occurrences of developmental features. Moreover, it has been shown that the primary vortex ring grows in both size and circulation until the (dimensionless) vortex formation time reaches approximately 4.[10] This critical value has also been associated with maximum thrust efficiency, inspiring further research that examined the optimality in biological propulsion and cardiovascular flow.[14,15]

In this paper, we present a bent-straw system that provides the framework for exploring fundamental concepts in vortex dynamics. This accessible table-top setup allows students to observe and measure characteristic features of vortices. Namely, we show how students can study the following properties of the vortex ring: (1) self-induced velocity of the vortex ring, (2) growth limit of the vortex ring, (3) interaction of the vortex ring with the wall, and (4) qualitative experiments of noncircular jet exit geometry on vortex ring propagation. The objective of these experiments is to hone students' intuition by visualizing core phenomena in vortex dynamics and exploring their theoretical basis and sometimes unintuitive results through experimentation. The swath of experiments suggested here covers the key characteristics of vortex dynamics that often escape undergraduate curricula.

## II. METHODS

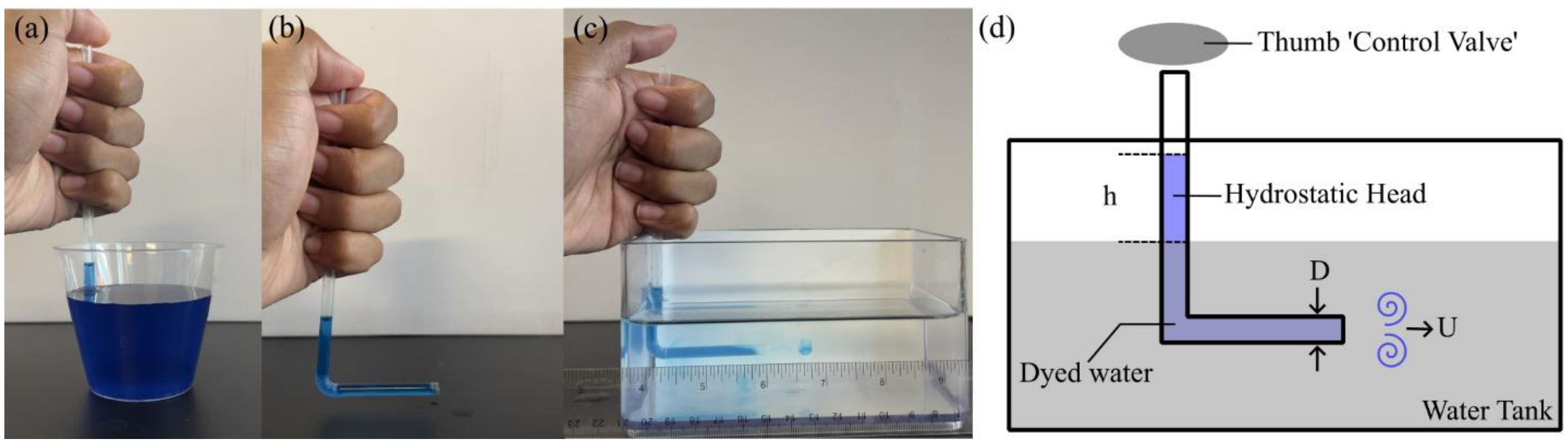

**Fig 1.** (Color online) Experimental workflow. (a) Dip the bent straw into the reservoir of dyed water and cover the free end with the thumb to (b) carry the dyed water in the straw. (c) Place the straw into the reservoir of clear water and release the thumb to generate a vortex ring. Note the ruler is used for length calibration during video analysis; using a second hand to stabilize the straw in the tank is recommended. (d) Schematic of bent-straw vortex ring generator annotated to show functional parallels between setups.

The experimental workflow is given in Figs. 1(a)-1(c): First, dipping the bent straw into the dyed water (colored with a water-based, food-grade dye), allows for a column of water to fill the straw. We recommend using a plastic bendable straw, which has an angle adjustable bellow. Covering the free end of the straw with a finger (thumb recommended) allows for transport of dyed fluid. The straw is then placed into the tank with clear water and should be stabilized against the wall of the tank with a free hand. Once the straw is in the water, hydrostatic head, h, can be measured. While h can be adjusted by changing the depth of the submerged straw, the straw nozzle should be sufficiently away from the water surface to avoid interaction with the water surface. Releasing the finger from the free end allows hydrostatic pressure to drive the dyed fluid out of the straw and cause vortex formation. A ruler should be laid against the tank such that it will be in-frame when recording. This process is captured with a video camera operating at 30-60 frames per second (fps), which is achieved by most smartphone cameras. Note, the specific fps used during videography must be recorded for future analysis. To demonstrate the effect of the nozzle geometry and the customizability of the experiment, vortices are created using a triangular nozzle by pinching three points along the circumference of the exit to create the vertices of a triangle. A straw with a large diameter (~1 cm) is recommended to properly create this phenomenon.

Adequate flow visualization is necessary but can sometimes be difficult to achieve. Although ring propagation is well captured with ambient lighting, additional lighting can be useful. Lighting should be carefully placed such that the ring passing through the lighting plane coincides with the camera viewing plane. A backdrop of any material, like vinyl tape, can also be used to increase contrast and highlight the ring.

Figure 1(d) shows the functional parallels between vortex ring generators in foundational experiments and our bent-straw system.[10-12] The notable analogues are the thumb covering the free end of the straw acting as the control valve, and the water column above the surface level acting as a *hydrostatic head* that drives ring formation. Although not drawn, a smartphone camera and video analysis act as the flow meter, and the reservoir with dyed water and the ruler used for calibration are shown in Figs. 1(a) and 1(c). Ultimately, the objective of this experiment is to explore concepts of vortex ring generation and interactions by using the bent-straw system to recreate these phenomena and tools like the Biot-Savart law to understand the dynamics.

The distance traveled by the dyed fluid at flow velocity $U$ over the jet injection duration $T$ moving through the straw outlet with diameter $D$ should always equal the hydrostatic head height of the dyed water column above the clear water (also termed the *stroke length*), $h$. Vortex formation time then reduces to

$$\hat{T} = \frac{h}{D}$$

which can be controlled by students. This form for $\hat{T}$ can be referred to as the *stroke ratio*, which describes the length of fluid ejected from the nozzle. The experimental advantage of vortex formation time is that it can be easily approximated by measuring the generation conditions (nozzle diameter and hydrostatic head height) but informs the experimentalist of important flow properties such as circulation, ring size and propagation velocity.

Video analysis to measure the change in hydrostatic head height can be completed in ImageJ v1.6.0, a free Java-based image processing program. The ‘reslice’ tool reconstructs a single 2D image from every frame in a video: a line is drawn across the region of interest, for example, the line shown in Fig. 2(a), and is ‘resliced’ such that the pixels along that line at each frame of the video are stacked together to create a single image that encapsulates temporal changes along that line. That is, at each frame, a snapshot of the hydrostatic head height is taken at the reslice line drawn in Fig. 2(a), and is stacked together to form the curve seen in Fig. 2(b). As the hydrostatic head height decreases along the reslice line, Fig. 2(b) shows how the hydrostatic head changes as the dyed water ejects through the nozzle exit. The slope of the resulting image can be measured to find $U$ in pixels per frame using $\tan(\theta)$, which can then be converted into more convenient units using the calibration target and recording fps.

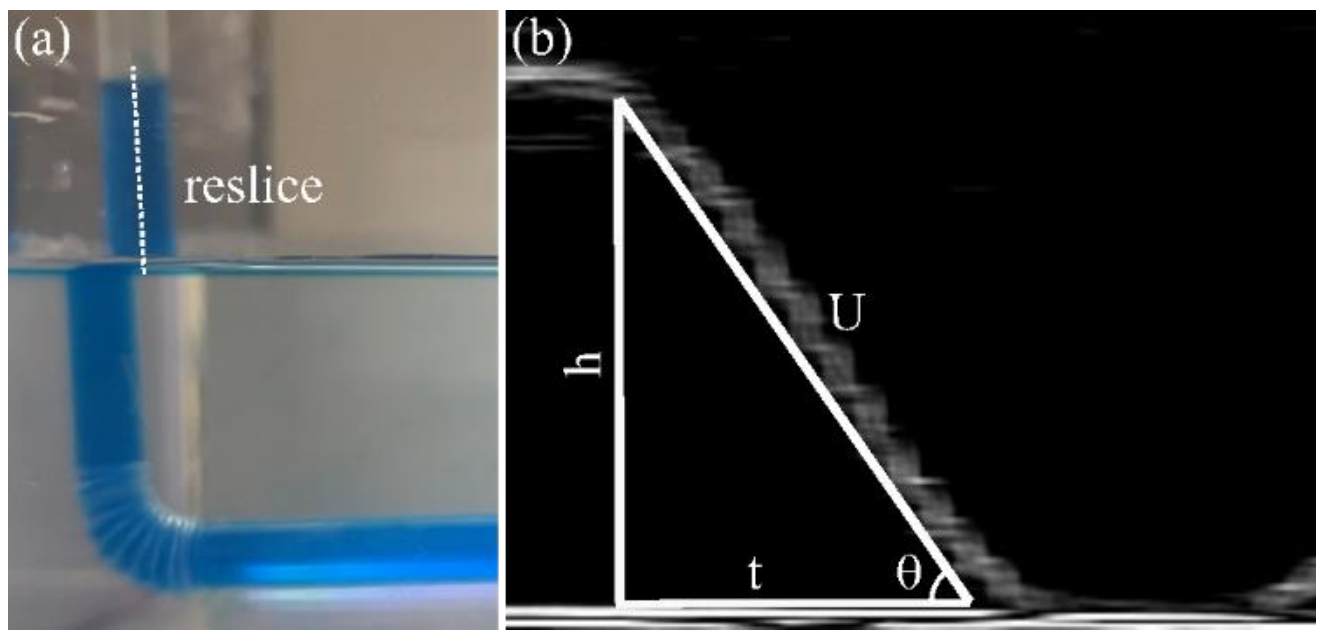

**Fig 2.** (Color online) (a) Reference image to illustrate where line for reslice is drawn. (b) Resliced image result. Hydrostatic head, h, (in pixels) moves down as a function of time, t, (in frames). The slope of the resulting line represents the flow velocity $U$, which can be found by using $\tan(\theta)$. Note that $\tan(\theta)$ approximates velocity $U$ in pixels per frame and must be calibrated to find $U$ in other physical dimensions.

## III. RESULTS AND DISCUSSION

### A. Self-induced propagation speed

One of the key properties of any vortex ring is its self-sustained propagation velocity that is different from the jet velocity. The speed of the vortex ring is determined by the strength of the vortex, i.e., its circulation. The self-induced velocity scales as $U_{vortex} \propto \Gamma/R$, where $\Gamma$ is the circulation and $R$ is the radius of the vortex ring. This is a direct result of the Biot-Savart law, which relates a vorticity field to its velocity field. While $R$ can be approximated from the imaged vortex ring, it is difficult to measure $\Gamma$ outright. One way to approximate the circulation is by integrating the vorticity flux across the exit (assuming all ejected fluid contributes to growth of the vortex ring's circulation). The circulation growth rate can be expressed as the convective flux of vorticity out of the pipe:

$$\frac{d\Gamma}{\mathrm{dt}} = \int \boldsymbol{\omega}(\boldsymbol{u} \cdot \boldsymbol{n})dA \tag{3}$$

In cylindrical coordinates, the largest component of vorticity generated from the fluid ejected at the nozzle is the azimuthal component, which can be expressed as $\omega_\theta = -\frac{dU}{dr}$. Substituting that component of the vorticity into Eq. (3) simplifies the circulation growth rate to

$$\frac{d\Gamma}{dt} = -2\pi \int U \frac{dU}{dr} dr = -2\pi \int \frac{d\left(\frac{U^2}{2}\right)}{dr} dr = \pi U^2$$

Further assuming the fluid ejection velocity is constant in time (i.e., constant slope in Fig. 2(b)), the total circulation accumulated over the injection time $T$ is then $\Gamma = \pi U^2 T$. Approximating $T$ as $h/U$, it follows that $\Gamma \propto Uh$. Consequently,

$$U_{vortex} \propto \frac{Uh}{R}$$

With this expression, students can vary the hydrostatic head height to see how it changes the propagation speed of the vortex ring. The propagation speed can be approximated by tracking the core with a reslice of its trajectory (or manually tracking its position at several frames using the multi-point tool in ImageJ). The jet velocity also changes with $h$ (Fig. 1(d)); students will need to measure the velocity of the jet. The radius of the ring is likely constant; however, this should be verified from imaging.

The learning objectives for this phenomenon are to visualize the vortex ring and understand its self-induced propagation speed. As motivation, students can illustrate the structure they expect to emerge from the exit before adding dye to the water—the visualized vortex ring might itself be unintuitive. When dyed and imaged, the ring propagation can also serve to visualize vorticity generation, which is the key feature of the ring structure and its dynamics. Students might also expect the ring to propagate at a similar velocity to the jet speed. However, the vortex ring propagates at a slower speed (Fig. 3). The reduced speed is caused by *entrainment*, the engulfing of nearby fluid into the vortex ring. Throughout propagation, the ring's composition changes as it contains fluid originally ejected from the nozzle and that which was entrained. Because any added fluid must be accelerated, this mixture continuously decreases propagation speed.

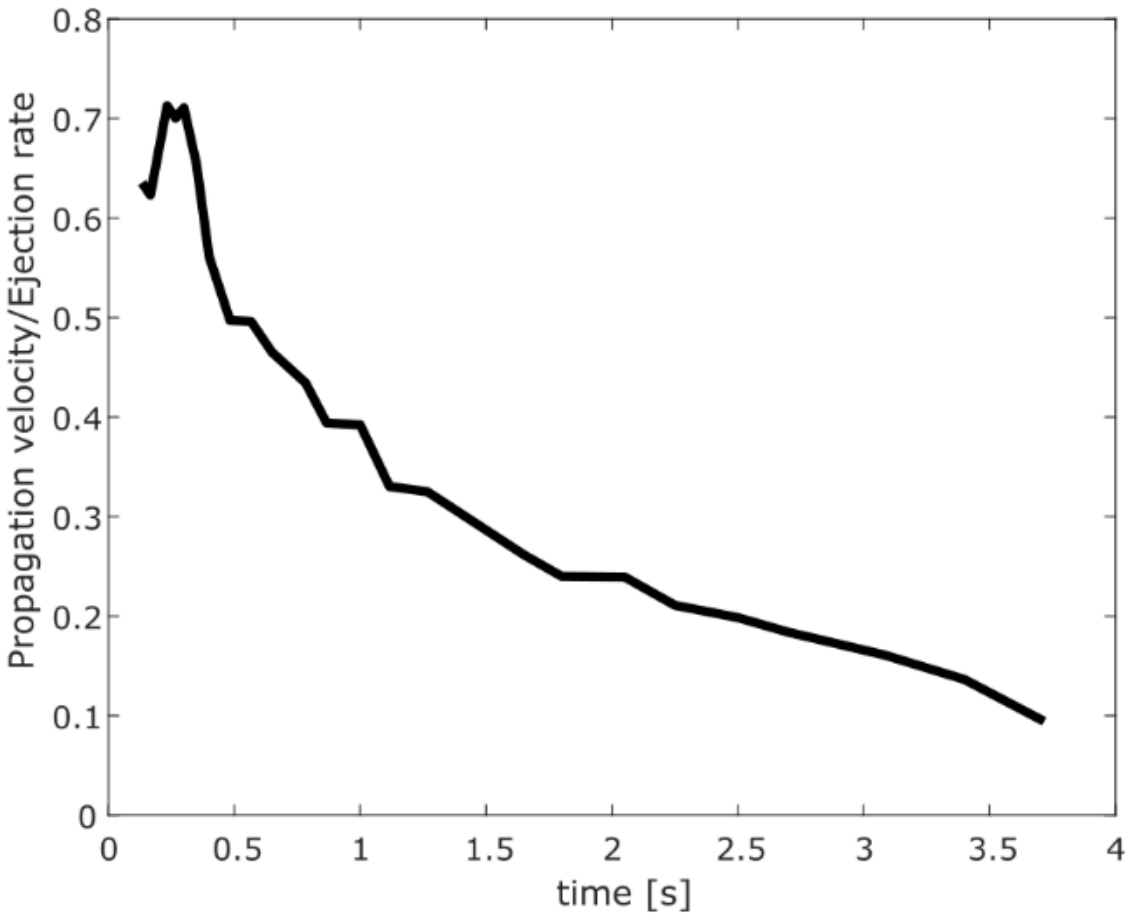

**Fig 3.** Ratio of vortex ring propagation velocity to rate of hydrostatic head height change (ejection velocity) as a function of time. The decrease in propagation velocity is caused by entrainment of surrounding fluid; momentum is continuously lost as entrained fluid is accelerated.

## B. The Limit of Vortex Ring Circulation

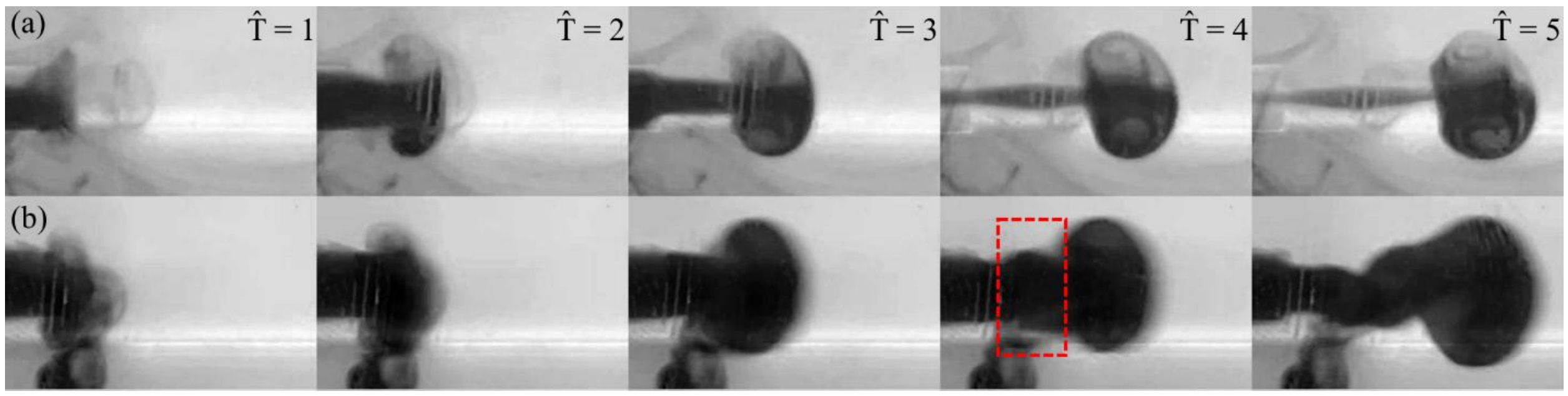

**Fig 4.** (Color online) Vortex ring with (a) stroke ratio of 3 and (b) stroke ratio of 8, imaged at vortex formation time equal to 1–5 from left to right, respectively. The first appearance of a secondary structure is highlighted with the red box.

Here, the learning objectives are to visualize secondary structures in the vortex ring and understand the limit of vortex ring growth as described by dimensionless vortex formation time. Vortices have intrinsic limitations that do not allow them to grow indefinitely. After a certain amount of vorticity has been fed into the ring, it can no longer accommodate more, and its circulation is saturated. Additional vorticity reorganizes into trailing secondary vortices. Moreover, the critical point at which the ring ceases to grow occurs at $\hat{T}\sim 4$, meaning if more vorticity is ejected beyond $\hat{T}\sim 4$, it will not contribute to the growth of the primary vortex ring.

In the bent-straw system, this critical limit is determined by the ratio between the initial hydrostatic head height *h* and *D*, i.e., the stroke ratio. Figure 4 compares vortices generated with a stroke ratio of 3 (Fig. 4(a)) and 8 (Fig. 4(b)) using the bent-straw and imaged at sequential formation times to show the roll-up process and ring core growth. The generated vortex is allowed to evolve beyond the vorticity injection duration. The ring in Fig. 4(a), for example, ingests vorticity up to $\hat{T} = 3$ when the stroke ratio is met and the entire length of fluid is ejected. Examining the ring at further injection times does not show a secondary trailing vortex. In comparison, the vorticity ejected beyond T~4 shows trailing secondary vortex (red box in Fig. 4(b) and Fig. 5). The clear presence of a single vortex ring for $\hat{T} < 4$ and trailing vortices for $\hat{T} > 4$ (Fig. 4(b) and Fig. 5) successfully demonstrates the limitation in vortex growth, consistent with the findings in Gharib *et al*.[10]

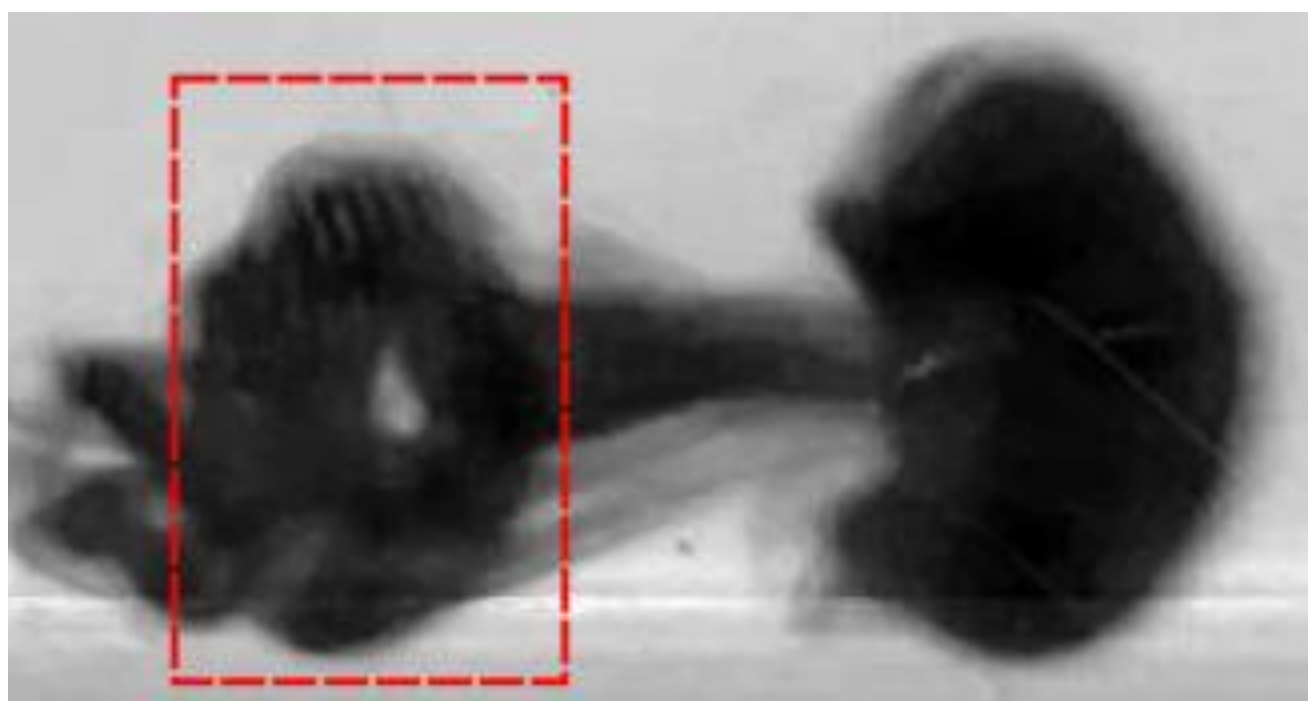

**Fig 5.** (Color online) Secondary vortex ring (red box) trailing behind the primary ring at $\hat{T} > 4$.

Follow-up discussions can further connect this experiment with other important concepts in fluid mechanics. On the topic of vortex dynamics, control volumes can be drawn at the nozzle exit to demonstrate Stokes' theorem and derive circulation; the ring structure can be used to demonstrate the vortex core and Helmholtz law; and the experimental setup can introduce other techniques like particle image velocimetry to quantitatively determine the flow field. Nondimensionalization is also an important topic as the theoretical basis for this experiment lies in nondimensional vortex formation time and circulation, which are both derived from circulation, propagation velocity, core diameter, and injection time.[10] The Buckingham Pi theorem and methods for nondimensionalization can be taught to derive these quantities, while the experiment helps relate the underlying physics of these numbers and underscores their meaning.

### C. Vortex Interactions with the wall (Image Vortices)

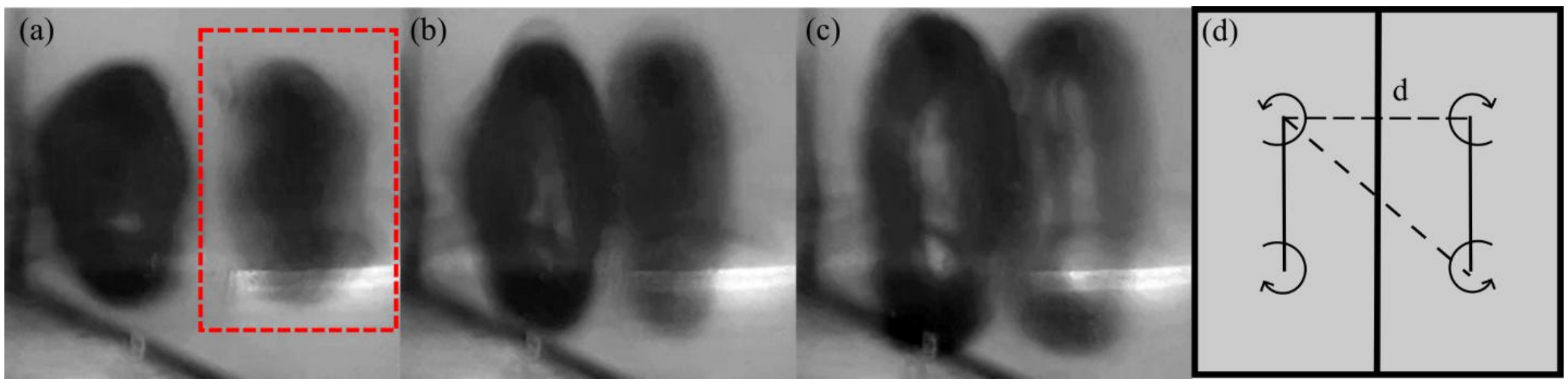

**Fig 6.** (Color online) Mirroring effect causing vortex ring expansion from (a) to (c) at $\hat{T} > 4$ with the mirrored vortex ring outlined in red. (d) Mirror effect modeled as two vortex rings approaching each other separated by a wall. Each vortex ring is modeled as two point vortices and some bi-directional interactions between the vortices, described by the Biot-Savart law, are drawn such that the dotted lines represent an anti-rotating induced velocity, and solid lines represent a co-rotating induced velocity.

Vortex ring's interaction with a wall presents another fascinating phenomenon. As the vortex ring approaches the wall, the lack of surrounding fluid for axial propagation forces radial expansion. This interaction can be seen by a horizontally propagating vortex ring interacting with a vertical wall or a vertically propagating vortex ring interacting with the floor by pointing the straw nozzle downwards at the bottom of the tank. Figure 6 shows the effect of the wall as the ring expands from frames (a) to (c).

This effect of the wall is also equivalent to that of two vortex rings approaching each other (except near the wall). Coincidentally, Fig. 6 shows the optically reflected image of the vortex ring as it approaches the wall. A simple 2D model can be used to demonstrate the physics of these two colliding vortices. For this, the kinematics of two counter rotating point vortices (a vortex with unit circulation concentrated at the core) can be solved analytically using the Biot-Savart law, which finds the velocity field can be calculated as $\mathbf{u}(x, y) = \Gamma/2\pi d\, \hat{\theta}$. Here, $d$ is the smallest perpendicular distance from the vortex to any point and $\hat{\theta}$ is the tangential unit vector. The vortex induces a velocity of magnitude $\Gamma/2\pi d$ at a point $(x, y)$. As one moves further away from the vortex, the induced velocity decreases.

Now, if there are two vortices of equal but opposite strength approaching each other, they each induce a counter-rotating velocity on each other, continuously decreasing their axial propagation speed as $d$ approaches 0. Each vortex also induces a co-rotating velocity radially, which causes the ring to expand. Each ring in Fig. 6 is represented by two point vortices and the interactions between each vortex are drawn in Fig. 6, where a solid line represents a bidirectional co-rotating effect and a dotted line represents a bidirectional counter-rotating effect. This can be solved analytically for the 2D vortex pair (see Supplementary Material).

For vortex ring interactions, the learning objectives are to gain understanding of the underlying physics of vortex ring interactions and to compare vortex-vortex and wall-vortex interactions, and to understand the mirroring effect. Students can track the trajectory of the vortex as it slows down and measure its expansion with the reslice tool. This can be compared with the 2D theory and students should find that the trajectories are initially similar but diverge as the ring approaches the wall due to viscous effects (i.e., the boundary layer formed at the wall; Fig. 7)

s

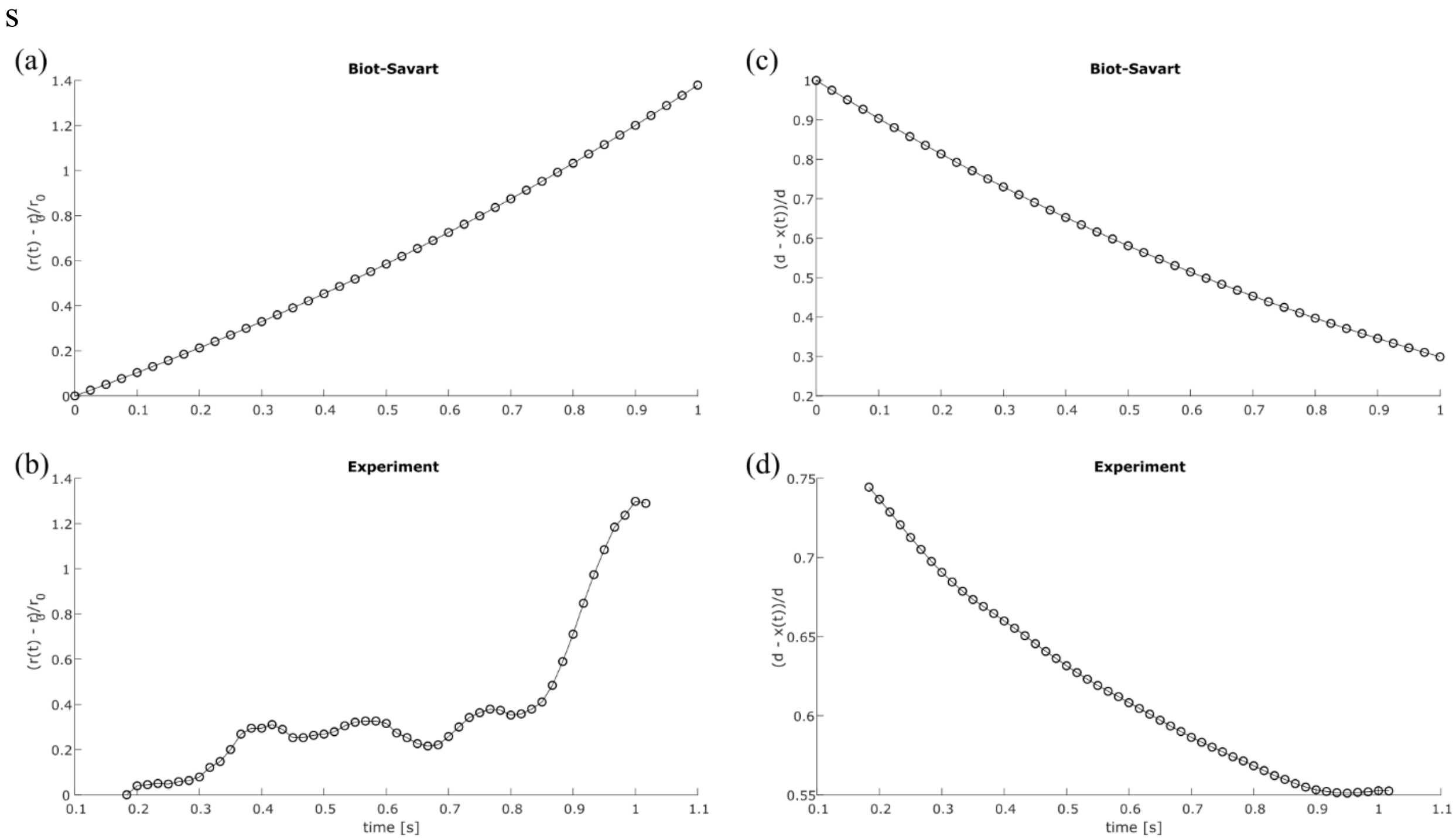

**Fig 7.** Kinematics of vortex mirroring effect. (a) and (c) radial expansion and axial propagation modeled by the Biot-Savart law for a 2D vortex pair, (b) and (d) experimental data of vortex ring approaching a wall.

## D. Non-circular Jet exit geometry

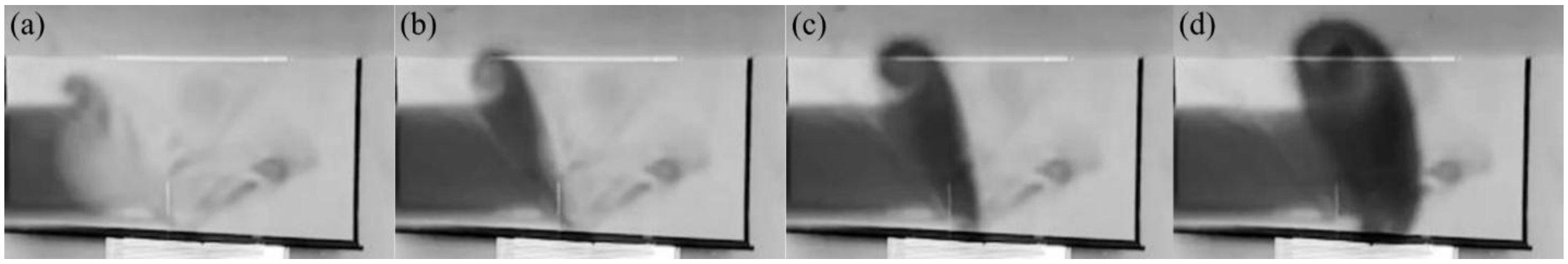

**Fig 8.** Vortex ring formation (a–d) in a 35-degree incline jet. The jet exit causes a ring core incline of 20 degrees.

Beyond exploration of stroke ratio in axisymmetric vortex ring formation, nozzle geometry can also be modified in the bent-straw system to reveal the effect of asymmetric jet exits on vortex formation. The complexity of flow at asymmetric jet exits prevents presenting simple solutions as above. Alternatively, these phenomena can be used to illustrate interesting phenomena in vortex dynamics or further discussions about governing physics.

Non-planar jet exits, for example, inclined nozzles, have been extensively studied as they can induce secondary structures and increase entrainment.[11,16] This geometry changes vortex ring dynamics by forcing asymmetric flow separation in the trailing edge. Longmire and Webster , for example, examined this geometry to understand its influence on ring formation, propagation, and entrainment and found that the near-exit vortex ring was altered to form at an angle slightly smaller than the incline angle.[11] The bent-straw system also replicates this result as the straw was cut with scissors at a 35-degree incline, while the ring exhibits a 20-degree incline (Fig. 8). As the high-pressure fluid approaches the exit plane, the asymmetric nozzle causes premature interactions with the low-pressure ambient fluid along the circumference causing circumferential variation in streamlines.[11] The Helmholtz law can then be used to describe local vortex tube stretching. Different planar cross-sectional geometries can also be used to alter vortex dynamics.

For asymmetric structures, a tighter radius of curvature along the vortex line will have a stronger inductive effect, which causes uneven forcing over the entire vortex ring resulting in bending or inversion. This inversion is observed in the vortex ring created with the bent straw with a triangular nozzle (Fig. 9). Here, the circular straw exit is easily folded down to form the triangular nozzle.

Much like the vortex emerging from the inclined jet, the triangular vortex ring enhances entrainment and finds its usefulness in mixing fluid. Miller conducted numerical simulations of elliptical, square, and triangular jet exits to examine the impact on entrainment and found triangular jets saw more than 100% increase in mass flow rate, where circular jets only see a 50% increase. For educational purposes, the relationship between the Biot-Savart law and fluid entrainment in this experiment can be emphasized: the triangular nozzle geometry creates corner flows in the vortex ring that will induce velocities in surrounding fluid, explained by the Biot-Savart law, which supports an increased mass flow into the ring and its inversion downstream.[15]

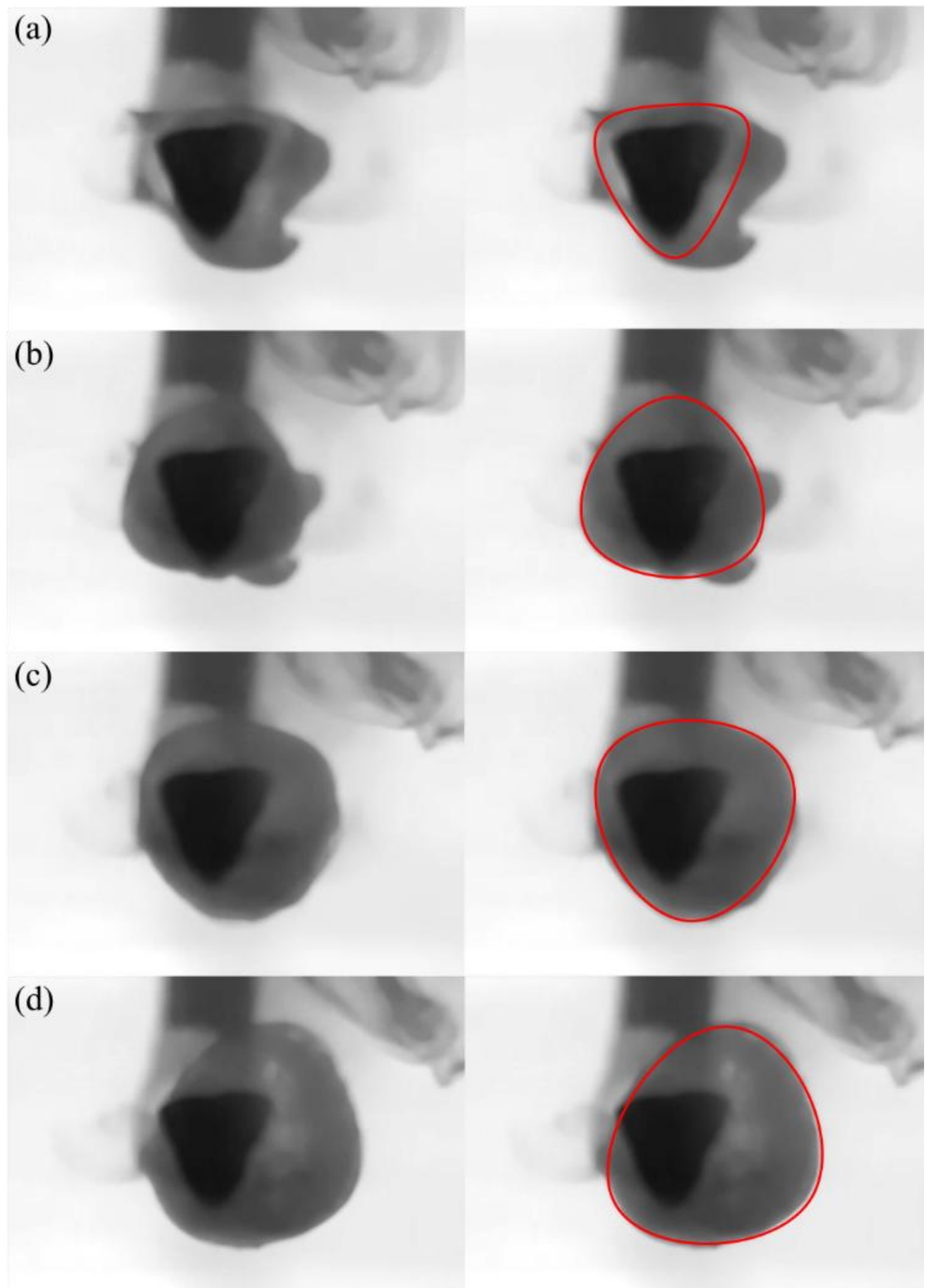

**Fig 9.** (color online) Propagation of a vortex ring (a–d) created with a triangular jet exit (left) with the ring outlined in red (right).

## IV. CONCLUSION

An affordable, student-safe, and controllable vortex ring generator using a bent-straw is introduced. We believe accessible demonstrations provide an entry point to fluids-relevant topics not ordinarily considered in undergraduate curricula. Ultimately, these tools create opportunities to actively engage students across different instructional levels. The bent-straw setup allows students to control key parameters in ring generation, visualize flow features; quantitatively and qualitatively explore the governing physics of vortex ring generation, its limiting processes, and interactions with other structures. Further discussion can also incorporate other concepts like control volume analysis and nondimensionalization techniques. The swirling motion of fluids is a fundamental component of its dynamics and its pedagogical introduction in undergraduate classrooms can be addressed with a simple tool. Beyond teaching students, the bent-straw vortex generator is meant to empower educators to incorporate this topic and emphasize its importance despite the challenges. The authors hope to see vortex dynamics and its long history of physics find a place in more classrooms with these tools.

## SUPPLEMENTARY MATERIAL

Please click on this link to access the supplementary material, which includes the MATLAB code for simulating the mirror effect for 2D vortex pair as shown in Fig. 7. Print readers can see the supplementary material at [DOI to be inserted by AIPP].

## ACKNOWLEDGMENT

We acknowledge the enthusiastic participation and contributions of Hannah Ceisler, Josh Martin, and Frank Fang during experiments in the bio-fluid mechanics course. E.J. was supported by the Colman Leadership Program at Cornell University. J.S. was supported by NSF-DGE 1922551 and NSF-DGE 2139899. C.R. gratefully acknowledges the funding from NSF-CMMI-204274, NSF-CBET-2442036, USDA-NIFA, Hatch grant accession #7003645, and Center for Teaching Innovation at Cornell University.

## AUTHOR DECLARATIONS

### Conflict of Interest

The authors report there are no conflicting interests to declare in conducting or reporting this research.

### Author Contributions

Elijah James: manuscript writing, investigation and development of experimental technique; Yukun Sun, Yicong Fu, Jena Shields, Cade Sbrocco: development of experimental technique; Christopher Dougherty: development of experimental technique, manuscript review and editing; Chris Roh: method conceptualization, investigation, development of experimental technique, and manuscript writing.